\documentclass[apj]{emulateapj}

\newcommand{\ms}{m\,s$^{-1}$}

\shorttitle{V450~And~B}
\shortauthors{He{\l}miniak et al.}
\usepackage{natbib}
\usepackage{mathrsfs}
\usepackage{amsmath,amssymb}
\usepackage{color}

\begin{document}

\title{SEEDS direct imaging of the RV-detected companion to V450 Andromedae, \\
and characterization of the system.}


%%=======================================
%-AUTHOR LIST----

\author{
K.~G.~He\l miniak\altaffilmark{1,2}, 
M.~Kuzuhara\altaffilmark{3,4,5}, 
K.~Mede\altaffilmark{6}, 
T.~D.~Brandt\altaffilmark{7,8},
R.~Kandori\altaffilmark{4}, %CA
T.~Suenaga\altaffilmark{4,9}, %CA
N.~Kusakabe\altaffilmark{5}, %obs, CA
N.~Narita\altaffilmark{5,4,9,6}, %LOI,comments
J.~C.~Carson\altaffilmark{10}, %comments
T.~Currie\altaffilmark{1}, %comments
T.~Kudo\altaffilmark{1}, %obs?
J.~Hashimoto\altaffilmark{5}, %obs?
L.~Abe\altaffilmark{11}, 
E.~Akiyama\altaffilmark{4}, %LOI
W.~Brandner\altaffilmark{12}, 
%S.~Egner\altaffilmark{1}, 
M.~Feldt\altaffilmark{12},
M.~Goto\altaffilmark{13},
C.~A.~Grady\altaffilmark{14,15,16}, %small comment
O.~Guyon\altaffilmark{1,17},
Y.~Hayano\altaffilmark{1},
M.~Hayashi\altaffilmark{4},
S.~S.~Hayashi\altaffilmark{1,9},
T.~Henning\altaffilmark{12},
K.~W.~Hodapp\altaffilmark{18},
M.~Ishii\altaffilmark{4},
M.~Iye\altaffilmark{4},
M.~Janson\altaffilmark{19},
G.~R.~Knapp\altaffilmark{20},
J.~Kwon\altaffilmark{6},
T.~Matsuo\altaffilmark{21},
M.~W.~McElwain\altaffilmark{14},
S.~Miyama\altaffilmark{22},
J.-I.~Morino\altaffilmark{4},
A.~Moro-Martin\altaffilmark{12,23},
T.~Nishimura\altaffilmark{1},
T.~Ryu\altaffilmark{9,4},
T.-S.~Pyo\altaffilmark{1},
%B.~Sato\altaffilmark{3},%other contribution?
E.~Serabyn\altaffilmark{24},
H.~Suto\altaffilmark{4,5},
R.~Suzuki\altaffilmark{4},
Y.~H.~Takahashi\altaffilmark{6,4},
M.~Takami\altaffilmark{25},
N.~Takato\altaffilmark{1},
H.~Terada\altaffilmark{4},
C.~Thalmann\altaffilmark{26},
%D.~Tomono\altaffilmark{1}, %???
E.~L.~Turner\altaffilmark{20,27},
M.~Watanabe\altaffilmark{28},
J.~Wisniewski\altaffilmark{29},
T.~Yamada\altaffilmark{30},
H.~Takami\altaffilmark{4},
T.~Usuda\altaffilmark{4},
and M.~Tamura\altaffilmark{6,4,5}
}
\email{xysiek@naoj.org}

%% are identified by the \altaffilmark after each name. Specify alternate

\altaffiltext{1}{Subaru Telescope, National Astronomical Observatory of Japan, 650 N.~A'ohoku Place, Hilo, HI 96720, USA}
\altaffiltext{2}{Subaru Fellow}
\altaffiltext{3}{Department of Earth and Planetary Sciences, Tokyo Institute of Technology, Ookayama, Meguro-ku, Tokyo, 152-8551, Japan}
\altaffiltext{4}{National Astronomical Observatory of Japan, 2-21-1, Osawa, Mitaka, Tokyo, 181-8588, Japan}
\altaffiltext{5}{Astrobiology Center of NINS, 2-21-1 Osawa, Mitaka, Tokyo, 181-8588, Japan}
\altaffiltext{6}{Department of Astronomy, University of Tokyo, 7-3-1, Hongo, Bunkyo-ku, Tokyo, 113-0033, Japan}
\altaffiltext{7}{Astrophysics Department, Institute for Advanced Study, Princeton, NJ, USA}
\altaffiltext{8}{NASA Sagan Fellow}
\altaffiltext{9}{Department of Astronomical Science, SOKENDAI (The Graduate University for Advanced Studies), 2-21-1 Osawa, Mitaka, Tokyo, 181-8588, Japan}
\altaffiltext{10}{Department of Physics and Astronomy, College of Charleston, 66 George St., Charleston, SC 29424, USA}
\altaffiltext{11}{Laboratoire Lagrange (UMR 7293), Universit\'e de Nice-Sophia Antipolis, CNRS, Observatoire de la C\^{o}te d'Azur, 28 Avenue Valrose, F-06108 Nice Cedex 2, France}
\altaffiltext{12}{Max Planck Institute for Astronomy, K\"{o}nigstuhl 17, 69117 Heidelberg, Germany}
\altaffiltext{13}{Universit\"{a}ts-Sternwarte M\"{u}nchen, Ludwig-Maximilians-Universit\"{a}t, Scheinerstr. 1, 81679 M\"{u}nchen,Germany}
\altaffiltext{14}{Exoplanets and Stellar Astrophysics Laboratory, Code 667, Goddard Space Flight Center, Greenbelt, MD 20771, USA}
\altaffiltext{15}{Eureka Scientific, 2452 Delmer, Suite 100, Oakland, CA 96002, USA}
\altaffiltext{16}{Goddard Center for Astrobiology}
\altaffiltext{17}{Steward Observatory, University of Arizona, 933 N Cherry Ave., Tucson, AZ 85721, USA}
\altaffiltext{18}{Institute for Astronomy, University of Hawaii, 640 N.~A'ohoku Place, Hilo, HI 96720, USA}
\altaffiltext{19}{Department of Astronomy, Stockholm University, AlbaNova University Center, SE-10691, Stockholm, Sweden}
\altaffiltext{20}{Department of Astrophysical Science, Princeton University, Peyton Hall, Ivy Lane, Princeton, NJ 08544, USA}
\altaffiltext{21}{Department of Earth and Space Science, Graduate School of Science, Osaka University, 1-1 Machikaneyamacho, Toyonaka, Osaka, 560-0043, Japan}
\altaffiltext{22}{Hiroshima University, 1-3-2, Kagamiyama, Higashihiroshima, Hiroshima, 739-8511, Japan}
\altaffiltext{23}{Department of Astrophysics, CAB-CSIC/INTA, 28850 Torrej\'{o}n de Ardoz, Madrid, Spain}
\altaffiltext{24}{Jet Propulsion Laboratory, California Institute of Technology, Pasadena, CA, 171-113, USA}
\altaffiltext{25}{Institute of Astronomy and Astrophysics, Academia Sinica, P.O. Box 23-141, Taipei 10617, Taiwan}
\altaffiltext{26}{Swiss Federal Institute of Technology (ETH Zurich), Institute for Astronomy,Wolfgang-Pauli-Strasse 27, CH-8093 Zurich, Switzerland}
\altaffiltext{27}{Kavli Institute for Physics and Mathematics of the Universe, The University of Tokyo, 5-1-5, Kashiwanoha, Kashiwa, Chiba, 277-8568, Japan}
\altaffiltext{28}{Department of Cosmosciences, Hokkaido University, Kita-ku, Sapporo, Hokkaido, 060-0810, Japan}
\altaffiltext{29}{H.~L.~Dodge Department of Physics \& Astronomy, University of Oklahoma, 440 W Brooks St Norman, OK 73019, USA}
\altaffiltext{30}{Astronomical Institute, Tohoku University, Aoba-ku, Sendai, Miyagi, 980-8578, Japan}
%-end of AUTHOR LIST
%%===========================================

\begin{abstract}
We report the direct imaging detection of a low-mass companion to a young, 
moderately active star V450~And, that was previously identified with the radial
velocity method. The companion was found in high-contrast images obtained with
the Subaru Telescope equipped with the HiCIAO camera and AO188 adaptive optics system.
From the public ELODIE and SOPHIE archives we extracted available high-resolution
spectra and radial velocity (RV) measurements, along with RVs from the Lick
planet search program. We combined our multi-epoch astrometry with these archival, 
partially unpublished RVs, and found that the companion is a low-mass star, not 
a brown dwarf, as previously suggested. We found the best-fitting dynamical masses to be
$m_1=1.141_{-0.091}^{+0.037}$ and $m_2=0.279^{+0.023}_{-0.020}$~M$_\odot$.
We also performed spectral analysis of the SOPHIE spectra with the iSpec code.
The {\em Hipparcos} time-series photometry shows a periodicity of 
$P=5.743$~d, which is also seen in SOPHIE spectra as an RV modulation of 
the star A. We interpret it as being caused by spots on the stellar surface, 
and the star to be rotating with the given period. From the rotation and
level of activity, we found that the system is $380^{+220}_{-100}$~Myr old, 
consistent with an isochrone analysis ($220^{+2120}_{-90}$~Myr).
This work may serve as a test case for future studies of low-mass stars, brown dwarfs and 
exoplanets by combination of RV and direct imaging data.

\end{abstract}


\section{Introduction}

For nearly two decades the radial velocity (RV) technique was the most effective one
in discovering extrasolar planets and brown dwarfs, as it was among the first 
techniques capable of detecting sub-stellar-mass objects orbiting normal stars.
Regular surveys began in the 1980-s \citep{fis14} and brought first results relatively early
\citep{lat89,may95}, although these were initially limited to relatively short period orbits,
owing primarily to the time span of the observations. Outside of indirect detections,
the direct imaging technique 
has been successful in exoplanet detection for over a decade now \citep{cha04},
but is still limited to objects relatively distant from the parent star,
even despite a tremendous improvement in recent years that allowed us to discover
less massive planets on closer orbits \citep{mar08,lag09,ram13,cur14}. 
Only very recently, thanks to the long-time coverage of RV data, and development 
of new generation instruments for both RV measurements and high contrast imaging 
at narrow separations, have the discovery 
spaces started to overlap, allowing for some objects to be detectable by both techniques.
This opens new possibilities in characterization of extrasolar planets, brown
dwarfs and low-mass stars. Only a few examples of RV and imaging detections of
a companion have been known to date, \citep[i.e.][also: J. Hagelberg et al. 2016, in preparation]{cre12a,cre12b,cre13a,cre13b,cre14,ryu16}
but their characterisations are either inadequate or uncertain due to incomplete 
data coverage of the orbit, or because the uncertainties in distance and mass of 
the primary were not adequately included in the final error budget. 
A notable case is, however, HD~16760b, first reported with RV measurements 
by \citet{sat09} and \citet{bou09}, and later detected in Keck sparse aperture 
masking observations by \citet{eva12}.
Initial discoveries indicated a minimum mass at the border between the planetary 
and brown dwarf regime: 13.13(56) and 14.3(9)~M$_J$ for \citeauthor{sat09} and
\citeauthor{bou09}, respectively. The Keck data revealed a low-inclination orbit, 
and a companion mass of 0.28(4)~M$_\odot$, but this result was not obtained from a 
simultaneous fit to RV and astrometric data, and the mass of the primary was assumed. 
It shows, however, how inaccurate the RV-based lower limits 
can be, and how important it is to supplement RV detections with astrometric data
\citep[see also][]{wil16}.

In this paper we present an example
where RV measurements span almost a whole orbital period, which allows us to 
obtain secure dynamical masses of both stars, and perform meaningful 
comparisons with the latest stellar evolution models. For the first time,
the orbital fit is obtained from all data simultaneously, and masses of both 
components are directly calculated. Basic information
about the target are given in Section \ref{sec_target}. Section \ref{sec_obsall}
describes the observational data used in this work. Orbital and spectral analysis
are described in Sections \ref{sec_orbSoln} and \ref{sec_spectr}, respectively.
In Section \ref{sec_disc} we discuss the activity and age of the system, and summarise 
our work in Section \ref{sec_sum}.

\section{The target}\label{sec_target}

\object{V450 And} \citep[HD~13507, HIP~10321, BD+39~496; $\alpha=02^h 12^m 55\fs0053$, $\delta=+40^\circ 40' 06\farcs0247$;][]{vLe07} is a known BY~Dra type variable. It is a G-type dwarf \citep{gra03}, located about 27 pc from the Sun \citep[$\varpi=37.25\pm0.55$~mas;][]{vLe07}, with a $V$-band apparent magnitude of 7.21 mag \citep{hog00}. It also forms a common-proper-motion pair with another BY~Dra type star \object{V451 And} (HD~13531, HIP~10339), and possibly belongs to the Castor moving group \citep{mon01,cab10}. 

The low-mass companion on a long-period orbit was first announced by 
\citet[][hereafter P03]{per03} in the Appendix of their paper. A preliminary
fit was performed on 19 ELODIE measurements spanning from January 1998 to
December 2002. P03 obtained an orbital period of $\simeq$3000 days, a small
eccentricity of 0.14, a companion minimum mass of 52~M$_{J}$, and a
projected major semi-axis of 4.3 AU, or 164 mas at the distance to the system,
assuming the mass of the host to be 1.09~M$_\odot$.
They had, however, poor coverage of the true orbital period, resulting in an underestimation of companion mass.
They also searched for the secondary with adaptive optics imaging, but failed
to identify it in their data, possibly because it was close to pericenter at that time. More such attempts were made later, 
but again without success, either because of the unfortunate location of the secondary
close to the primary \citep{met09}, or lack of sufficient sensitivity
\citep{tok14,rid15}. We present the first
positive detection of V450~And~B with high-contrast imaging.

\section{Observations and data}\label{sec_obsall}

\subsection{SEEDS Observations}
We observed V450 And as part of the Strategic Exploration of Exoplanets and Disks 
with Subaru \citep[SEEDS;][]{tam09} survey, which has searched for exoplanets and imaged circumstellar disks around hundreds of nearby stars \citep[e.g.,][]{tha11,has12,jan13}. 
One category of SEEDS targets consists of young nearby stars that can be age-dated using their rotation periods or chromospheric activities as clocks \citep[see][]{kuz13}, and V450~And is included in this category (we discuss the star's age in Sections \ref{sec_age} and \ref{sec_iso}). Its science case, however, also resembles the category of targets which are known to have inner planets or a long-term RV trend \citep{nar10,nar12,ryu16}. \par

We observed V450~And with the HiCIAO camera \citep{suz10}, a high-contrast imaging instrument on the Subaru Telescope. The adaptive optics system AO188 \citep{hay10} was used to reduce the image degradation caused by atmospheric turbulence and improve the Strehl ratio. 
AO188's atmospheric dispersion corrector \citep[ADC;][]{egn10} removed the chromaticity of atmospheric refraction.
None of our observations used a focal plane mask, but all were performed with a Lyot stop in the pupil plane.  HiCIAO's original Lyot stop was replaced in September 2013 with a larger, circular pupil stop (in other words, smaller pupil-plane mask) to increase throughput and improve angular resolution.

\begin{figure}
\centering\includegraphics[width=\columnwidth]{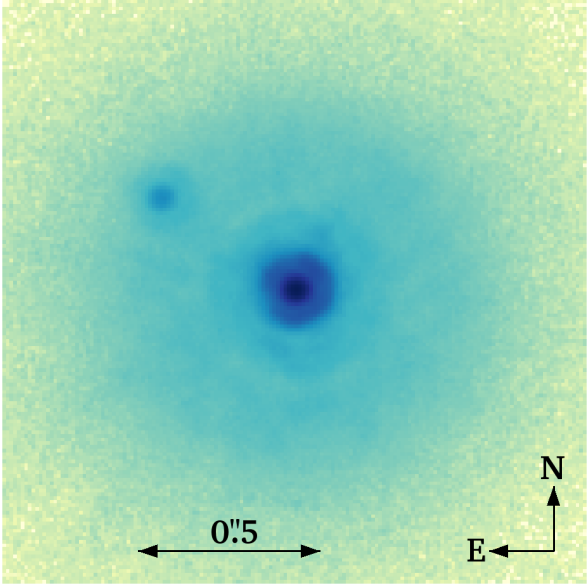}
\caption{$H$-band image of V450And showing its M dwarf companion about $0\farcs44$ to the northeast. The stretch is logarithmic; the companion is about 40 times fainter in the $H$-band than the G5 primary. The figure is a composite image of 50 single, unsaturated frames, all taken on 8 January 2015. The total exposure time is 575 sec.}
\label{fig_image}
\end{figure}

We observed the star at four epochs, controlling the image rotator to fix the pupil rotation relative to the camera (angular differential imaging, or ADI, mode, \citealt{mar06}). Each observing sequence consisted of longer sequences of frames in which the central star's point-spread function (PSF) was saturated, and shorter sequences using neutral density filters to avoid saturation. The unsaturated images were taken sparsely during one visit. The saturated images were used to search for faint companions (Kuzuhara et al.~2016,~in preparation) while the unsaturated images were used for image registration and flux calibration.  We discovered one companion candidate, about 40 times fainter than V450 And in the $H$-band; follow-up observations confirmed it as a low-mass stellar companion.  Figure \ref{fig_image} shows the pair V450 And AB. Because of the modest contrast between the primary and secondary, we used only unsaturated images in each observing sequence to measure the astrometry of the system and to track its orbital motion.  Table \ref{tab:obslog} summarizes the unsaturated data used in our analysis of the V450~And system; our analysis makes no use of saturated imaging.
The next section discusses the measurement and calibration of the astrometry in each observing sequence.  

\begin{deluxetable*}{cccccccc} 
\tablewidth{0pt}
\tablecaption{Observing Log for Unsaturated Data}

\tablehead{
    \colhead{Obs. Date} &
    \colhead{$N_{\rm{exp}}$ } &
    \colhead{$t_{\rm{tot}}$} &
    \colhead{Filter} & 
    \colhead{Mean Airmass} &
    \multicolumn{2}{c}{Field Rotation (deg)} &
    \colhead{$\rm{\Delta}$mag} \\
    \colhead{(UT)} &
    \colhead{} &
    \colhead{(s)} &
    \colhead{} & 
    \colhead{} &
    \multicolumn{1}{c}{Total} &
    \multicolumn{1}{c}{Average} &
    \colhead{}
    }

\startdata
2012 Nov 05 & 10 & 100 & $J$ & 1.09 & 1.0 & 0.12 & 4.23 $\pm$ 0.07 \\
-- & 25 & 125 & $H$ & 1.13 & 24 & 0.054 &4.10 $\pm$ 0.05\\
-- & 10 & 50 & $K_{\rm{s}}$ & 1.09 & 0.66 & 0.073 & 3.85 $\pm$ 0.07\\
2013 Jan 02 & 16 & 40 & $J$  & 1.39 & 0.96 &0.024 & 4.18 $\pm$ 0.08 \\
-- & 7 & 62 & $H$ & 1.43 & 0.60 & 0.044 & 3.95 $\pm$ 0.06\\
-- & 16 & 24  & $K_{\rm{s}}$ & 1.42 & 0.74 &0.020 & 3.82 $\pm$ 0.08\\
2013 Oct 16 & 13 & 195 & $H$ & 2.19 & 1.0 & 0.062 & 4.05 $\pm$ 0.04 \\
2015 Jan 08 & 50 & 575 & $H$ & 1.08 & 51 & 0.10 & 4.03 $\pm$ 0.03
\enddata
\tablenotetext{}{ Note: The data sets shown in this table were obtained without a focal-plane mask but with a Lyot stop in a pupil plane (see Section \ref{sec_obsall}). The average field rotation corresponds to the average of each rotation angle during an exposure. Sequences of unsaturated images were taken alternately with saturated frames.}

\label{tab:obslog}
\end{deluxetable*}

\subsection{Data Reductions and Astrometry} \label{reduction astrometry}

At an $H$-band contrast of only $\sim$40, V450~And~B is bright enough to be detected in the unsaturated short-exposure data sets. This enables us to simultaneously obtain very accurate positions and photometry of both stars: our simultaneous relative photometry is unaffected by temporal variations in the Strehl ratio.
We used the ACORNS pipeline \citep{bra13} to correct bad pixels, remove correlated read noise, and to correct the field distortion. We did not, however, apply any algorithms to subtract the primary star's PSF, apart from our removal of an azimuthally symmetric halo.  This ensures that there is almost no self-subtraction of the companion's PSF.  Our goal is to obtain precise relative astrometry and photometry, not to search for very faint companions.

We used only the $H$-band observations listed in Table \ref{tab:obslog} to measure relative astrometry, as HiCIAO's distortion correction is best-characterized in this band.
We compute the distortion correction of the HiCIAO camera using images of the dense cores of the globular clusters M5 and M15 \citep{bra13}. We then compare the HiCIAO images of M5 or M15 with the archival M5 or M15 images obtained by {\it Hubble Space Telescope} and Advanced Camera for Surveys (ACS). The distortion, plate scale, and true north of ACS have been well calibrated \citep[cf.][]{Anderson_2006_ACS, Anderson+King_2004_ACS, van_der_Marel_2007_ACS}. With good seeing, the fractional uncertainties on the distortion are $\sim$10$^{-4}$ near the center of the field. The nonlinear component of the distortion is extremely stable between observing runs, with nearly all of the variation confined to the plate scale. We therefore adopt our nonlinear distortion corrections from runs with good observing conditions. \par

In the case of 2012 November 5, the seeing was poor, so we used observations of M15 from that date to compute the plate scale and use the high-quality 2011 May distortion map for nonlinear-distortion correction. Likewise, for the 2013 January run, we corrected for only the plate scale based on the M5 data that were acquired in the same run, and the nonlinear distortions were corrected using the 2011 May distortion map. HiCIAO's Lyot stop was replaced before the 2013 October run to improve throughput and angular resolution, necessitating a new measurement of the distortion correction.
In the 2013 October and 2015 January runs, we observed globular clusters for the distortion corrections.  High-quality data of M5 are available for 2015 January run. Unfortunately, the observing conditions for the 2013 October M15 calibrations were too poor to permit a good measurement of the plate scale and the other distortion components. For this data set we instead used observations of M15 taken in 2013 November, and added an additional uncertainty of 0.25\% to the plate scale, corresponding to the measured run-to-run scatter. We find that the new Lyot stop, installed in September 2013, significantly affected HiCIAO's PSF but had a negligible impact on distortion correction.

After correcting field distortion, we measured the PSF centroids of the primary star on each individual frame by fitting elliptical Gaussians.   Then, all the frames from a given observing night were shifted to a common center.
Next, the radial profile of the primary star's PSF was subtracted from each data frame 
to remove the PSF halo.  Failing to remove the PSF halo would bias our measurements both of the companion's flux and position, as the halo's mean and gradient are both nonzero at the location of V450 And B.  We  injected artificial point sources to determine any loss of companion flux due to halo subtraction.  We applied these correction factors, though they are mostly very small ($\ll$10\%).  Spatial variations in the Strehl ratio are negligible at the $0\farcs5$ separation of the binary, while HiCIAO's flat-field images are stable to $\sim$2\% over a period of years \citep{bra13}, smaller than all of our derived photometric uncertainties.
Because our data were obtained with the image rotator maintaining the orientation of the PSF (ADI mode), the PSF-subtracted frames were de-rotated to align their $y$ axes to celestial north. Finally, the de-rotated frames were combined into a single final frame. We fit an elliptical Gaussian to measure the centroid of V450 And B in this final, halo-subtracted frame; Table \ref{tab:astrometry} lists the relative astrometry for each epoch. Figure \ref{fig_image} shows the final image for the observations at 2015-01-08 without subtracting the radial profile of V450 And A. Note that our analyses for the companion V450 And B are all based on the images with subtracted PSF halos.

To derive the flux ratio between the components, we compare the photometry of the central star and the companion in the same final, combined image. As described above, we measure the companion flux after subtracting the central star's halo and applying a small correction for flux loss. In order to estimate the error in our measured contrasts, the final image is first convolved with a circular aperture having a radius equal to the full width at half maximum (FWHM) of the PSF. As in \cite{bra13}, we use the convolved image and compute the standard deviation of residual signals in an annulus surrounding the central star at the same separation as the companion. This scatter is much larger than the companion's photon noise.

Our four data sets obtained over three nights contain a small number of pixels with significant ($>$3\%) nonlinearity. Removing frames with one or more pixels above this nonlinearity threshold has a negligible effect on our results, apart from a 0.04 mag difference in photometry for the 2013 January $H$-band observations.  We do not exclude these frames when deriving our final photometry (Table \ref{tab:obslog}) and astrometry (Table \ref{tab:astrometry}).

\subsection{Astrometry Error Analysis}

Our error budget for astrometry is dominated by errors in fitting elliptical Gaussians and in determining the plate scale, with lesser contributions from uncertainties in the nonlinear image distortion and from differential atmospheric refraction. We evaluate the uncertainties in the distortion correction using a Markov Chain Monte Carlo (MCMC) method to sample the space of coefficients in the distortion correction polynomial \citep[see][]{bra13}. Where good images of a dense star cluster are available (as for the 2015 January run), the resulting separation errors are negligible ($\lesssim$0.1 mas). For some runs, however, simultaneous calibration data are unavailable or are of poor quality. In these cases, we use the 2011 May calibration for the nonlinear distortion and for the orientation of true north. 
We estimate the resulting uncertainty using the scatter among other runs with good calibration data. The scatter due to the nonlinear distortion is negligible along the vertical axis of the detector, but ranges from $\lesssim$0.1 mas up to $\sim$0.3 mas along the horizontal axis depending on the location in the image where the star is observed. The scatter in the angle of true north is negligible, 0\fdg031 ($5 \times 10^{-4}$ rad).

For all but the 2013 October run, we measured the plate scale and its uncertainty using MCMC on simultaneous calibration images. The resulting uncertainties are very small, $\lesssim$0.16 mas at a separation 0\farcs5. The 2013 October run lacks simultaneous calibration data, so we used the full distortion correction measured in 2013 November. We note that a plate scale variation between each run contributes to an astrometric uncertainty. We estimate the resulting astrometric uncertainty from the plate scale to be a non-negligible $\sim$1.1 mas by computing the plate scale scatter among 15 other runs with identical instrument configurations (see above for an uncertainty due to the variation of nonlinear distortions).

Finally, we determine the centroid error by calculating the scatter of star-companion separations among sets of frames in an imaging sequence. Before calculating the scatter, all the individual PSF-subtracted frames for a run were divided into several groups and the data in each group were combined. We omitted this step for 2013 January 2 since we have just seven frames, and directly calculated the scatter of separations among those seven frames. The scatter in astrometry among groups of frames ranges from 0.03 to 0.12 pixels (0.3 to 1.1 mas). This is larger than the field distortion errors for all but the 2013 October run, for which simultaneous distortion data were unavailable. We do not independently incorporate an error of the primary star's centroid into the total error budget.  Instead this error is inserted into the astrometry error by computing a scatter of separation measurements between the primary and secondary. This flows from the fact that scatter of the primary star's centroids result in a scatter of measurements of the primary-secondary separations. 

We took all our data with the ADC \citep{egn10} in place. This set of prisms corrects the dependence of the PSF center on wavelength, removing the systematic shift in centroid due to difference in spectral type. Atmospheric refraction also shifts PSF centroids over the field-of-view. This effect is negligible ($<0.1$~mas) for all but the 2013 October observations, which were conducted at airmass 2.19. In that case, we computed the index of refraction as in \citet{mat04,mat07}, taking into account the weather conditions during observations: outside atmospheric pressure of 616.3~hPa, outside temperature of 273.19~K, and humidity of 46 percent, as given in the CFHT weather archive\footnote{http://mkwc.ifa.hawaii.edu/archive/wx/cfht/}. We then corrected the astrometry using a somewhat simplified version of the approach described in \citet{hel09}. A correction of $\sim$0.4~mas was applied to the relative separation, while $-$0\fdg008 was applied to the position angle. We have added these offsets to our astrometry, but note that they are smaller than the astrometric uncertainties. The error bar for each astrometry in Table \ref{tab:astrometry} includes the uncertainties of centroiding, plate scale, angle of true north, and non-linear distortions.

\begin{deluxetable}{cccc}
\tablewidth{0pt}
\tablecaption{Astrometric Measurements}

\tablehead{
    \colhead{Obs. Date} &
    \colhead{MJD} &
    \colhead{Separation Angle} &
    \colhead{Position Angle }  \\
    \colhead{(UT)} &
    \colhead{(days)} &
    \colhead{(mas)} &
    \colhead{(\arcdeg)} } 
    
\startdata
2012 Nov 05 & 56236.342 & 437.87 $\pm$ 0.38 & 73.095 $\pm$ 0.062 \\
2013 Jan 02 & 56294.375 & 437.5 $\pm$ 1.2 & 71.76 $\pm$ 0.17 \\
2013 Oct 16 & 56581.652 & 432.7\tablenotemark{a} $\pm$ 1.2 &  65.512\tablenotemark{a} $\pm$ 0.089 \\
2015 Jan 08 & 57030.244 & 422.93 $\pm$ 0.43 & 55.61  $\pm$  0.055
\enddata

\tablenotetext{a}{Corrected for the effect of atmospheric refraction. The separation and position angle before the correction are 432.3 mas and 65\fdg520, respectively.}

\label{tab:astrometry}
\end{deluxetable}    

\subsection{Archival Spectra and Radial Velocities}
The complete set of archival spectra and RV measurements consists of
80 observations coming from three instruments. The total time span of 
the data is over 18 years, which nearly covers the full orbital period.
We present them in Table \ref{tab_rvall} in the Appendix.

\subsubsection{ELODIE}
The ELODIE spectrograph is the decommissioned instrument that was attached to the
1.9-m telescope in the Observatoire de Haute-Provence (OHP) in France.
From the public ELODIE archive \citep{mou04}\footnote{http://atlas.obs-hp.fr/elodie/}
we extracted 25 RV measurements, including six previously unpublished by P03.
The total time span is 2536 days (1998 January 8 to 2004 December 18). With one exception,
the exposure time varied between 600 and 1200~s; for the majority of observations
it was 900~s. The signal-to-noise ratio (SNR) varied from 50 to 116. In one case 
the exposure time was only 90~s, with a corresponding SNR of 29. The majority of spectra were taken in the simultaneous object-calibration mode (OBTHs), designed for higher RV precision. Two spectra were taken in a different, less precise mode (OBJOs).
We took the RV measurements that are directly listed in the archive, and which are given with the precision of 10~\ms. We recalculated some 
of the RVs by ourselves (using cross-correlation technique) and found that this 
precision is sufficient in this case, especially taking into account the 
systematics and the jitter originating from the stellar activity (see further 
Sections). We also calculated the photon-noise RV errors using the formula 
from \citet{bar96}. Following \citet{sou13}, we added in quadrature a systematic 
RV uncertainty of 15~\ms\,for OBTHs observations, and 30~\ms\,for OBJOs.

\subsubsection{SOPHIE}
In the public SOPHIE archive\footnote{http://atlas.obs-hp.fr/sophie/} we found 
another 25 spectra, taken between 2013 September 26 and 2016 February 29 (time span: 886 days).
The SOPHIE spectrograph is the current instrument of the OHP 1.9-m telescope.
SOPHIE observations were done twice, sometimes three times per night, and the RVs are 
much more precise than that of the ELODIE data. It is notable that the archive lists
different values of velocities 
than are given in the available CCF headers. We have checked that the latter are more
reliable. We have also extracted values of the bisector span, as well as the processed
1D spectra, which we later used for spectral analysis. The exposure times varied from
600 (majority of the spectra) to 900 seconds, and the SNR from 70 to even 190 in one case.
Following \citet{sou13}, we added in quadrature a systematic RV uncertainty of 
4~\ms\, to all the photon-noise uncertainties, which we took directly from the archive.

\subsubsection{Hamilton}
Recently, \citet{fis14} published measurements of the Lick planet search program,
carried out using the Hamilton spectrograph at the 3.5-m Shane telescope in the 
Lick observatory in California. Available are 30 data points spanning 5012 days
(1998 January 18 to 2011 October 9), which is the longest time span of all the data 
sets we used. It also covers the gap between ELODIE and SOPHIE observations, which 
was crucial for merging all the data in the orbital fit. Contrary to 
SOPHIE and ELODIE, the Hamilton measurements are not given for the instrument's 
zero-point, but instead are shifted, so their average is 0. The SNR of the spectra 
varies from 60 to 200, in most cases exceeding 100. Exposure times are not given.

\subsection{{\em Hipparcos} Photometry}\label{sec_hip_phot}
V450~And is a BY~Dra-type variable, and the photometric variability is likely related 
to the presence of spots on the rotating disk of the star, so its period represents
the period of rotation. Unfortunately, the value of 7.6~d given in \citet{str00}
is marked as uncertain. A value of 7~d can also be found in the literature 
\citep{isa10}, but it was found from period-activity relations, not measured 
directly, and we also find it unreliable.

\begin{figure*}
\centering\includegraphics[width=0.7\textwidth]{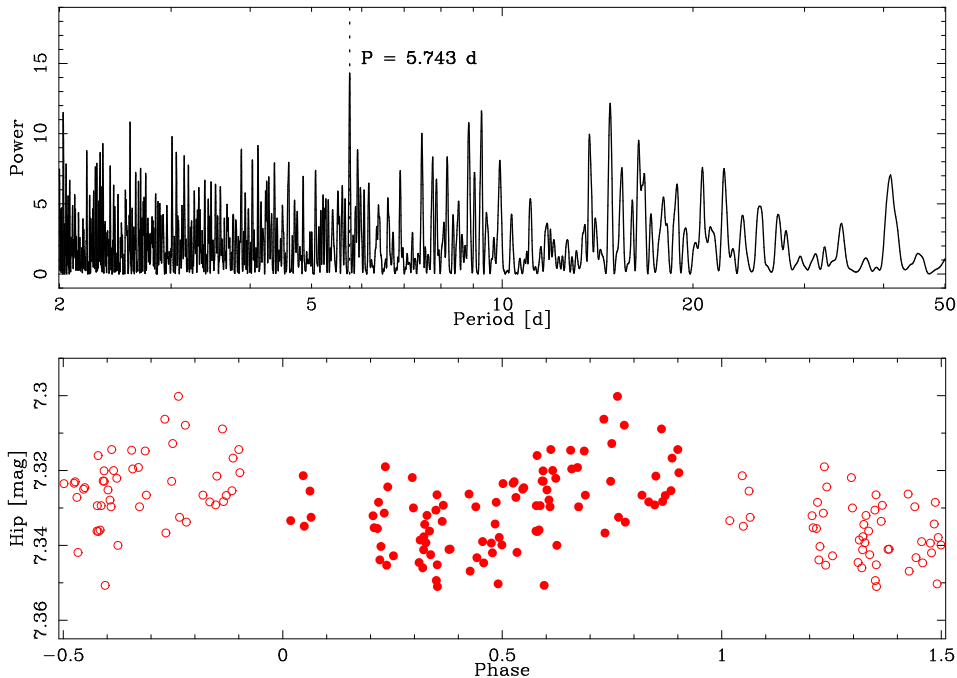}
\caption{{\it Top:} The Lomb-Scargle periodogram of the {\em Hipparcos} photometry,
with the most distinct peak at $P=5.743$~d marked. {\it Bottom:} The {\em Hipparcos} 
data phase-folded with the $5.743$~d period. For clearance, we replot the data with
open symbols for phases $<$0 and $>$1.
}\label{fig_per}
\end{figure*}

In order to assess the rotation period we took the {\em Hipparcos} time-series
photometry \citep{per97} and ran a Lomb-Scargle 
periodogram\footnote{Periodograms for this work were
created with the on-line NASA Exoplanet Archive Periodogram Service:\\ 
http://exoplanetarchive.ipac.caltech.edu/cgi-bin/Pgram/nph-pgram}. 
We focused on periods between 2 and 50 days, and set the constant step in frequency of 
$1/10W \simeq 9.13\times10^{-5}$~d$^{-1}$, where $W\simeq1096$~d is the time 
span of {\em Hipparcos} data. The most distinct peak is found at 
$P=5.743\pm0.003$~d or $f=0.17411(91)$~d$^{-1}$. The periodogram and
{\em Hipparcos} data phase-folded with this period are shown in Figure~\ref{fig_per}.
%5.74335 d

\section{Orbital solution}\label{sec_orbSoln}

Both the radial velocity and astrometry data were fit simultaneously with a Keplerian model to solve for the system's orbital parameters. The full solution includes 9 parameters, with a further 3 for the spectrograph offsets. To perform the fitting, we adopt a Bayesian approach similar to that discussed in \citet{ford2006}. The posterior probability distributions for the model parameters are proportional to the product of the model parameter's likelihood, $\mathscr{L}({\rm Model})$, and their prior probability based on previous knowledge, $p({\rm Model})$. Assuming the data errors ($\sigma$) are independent and follow a Gaussian distribution, the likelihood may be written as
\begin{equation}\label{eq:likelihood}
	\mathscr{L}({\rm Model}) = \exp\left(-\frac{\chi^2}{2}\right)
\end{equation}
with 
\begin{equation}\label{eq:chiSquared}
 \chi^2 = \sum_{i}\frac{({\rm Model}_i-{\rm Data}_i)^2}{\sigma^2_i} ,
\end{equation}
and the priors used in our model fitting are summarized in Table \ref{tab_priors}.

\begin{deluxetable}{cc}
\tablewidth{0pt}
\tablecaption{Adopted Bayesian priors}
\tablehead{
    \colhead{Parameter} &
    \colhead{Prior}}
\startdata
$\varpi$  & Gaussian\tablenotemark{a}$\times(1/\varpi^4)$ \\
$e$ & $p(e)\propto e$\tablenotemark{b}   \\
$P$  & Logarithmic \\
$i$  & $p(i)\propto \sin(i)$\\
$m_1$ \& $m_2$ & PDMF\tablenotemark{c} \\
others & Uniform 
\enddata
\tablenotetext{a}{{\it Hipparcos} value of $37.25 \pm 0.55$ mas, from \cite{vLe07}}
\tablenotetext{b}{From \citet{duq91}}
\tablenotetext{c}{Present-Day Mass Function \citep[][Table 1]{chab03}}
\label{tab_priors}
\end{deluxetable}

Our Keplerian model includes the directly varied parameters: $m_1$, $m_2$, $e$, $P$, $\varpi$, $T_0$, $i$, $\Omega$, $\omega$, $\gamma_1$, $\gamma_2$, $\gamma_3$; respectively these are, the mass of the primary star, mass of companion,  eccentricity, period, parallax, time of last periapsis, inclination, longitude of the ascending node, argument of periapsis, and the radial velocity offsets for each spectrograph (ELODIE, Hamilton and SOPHIE). Including the parallax as a varied parameter ensures its estimated errors and prior probability distributions are appropriately included into the posteriors. Fitting these 12 parameters with a direct sampling approach, such as simple Monte Carlo, can be difficult due to their possibly complicated likelihood topography. To overcome this we use the Markov Chain Monte Carlo (MCMC) features of a new software package entitled the Exoplanet Simple Orbit Fitting Toolbox (ExoSOFT; K. Mede \& T. Brandt 2016, in preparation). This toolbox explores the parameter space using a multi-stage approach ending in MCMC. It is capable of fitting any combination of astrometry and radial velocity data, and performs automated post-processing to summarize the results. The code was written primarily in the Python programming language with the computationally intensive model in C++.

\begin{figure}
\centering\includegraphics[width=\columnwidth]{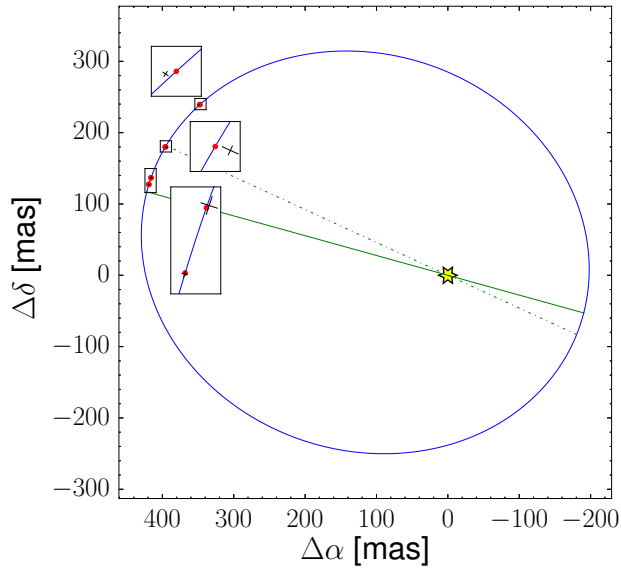}
\caption{Astrometric orbit of V450~And~B relative to A (marked as yellow star). The line of apsides and line of nodes are marked with the solid and dotted green lines, respectively. Our HiCIAO measurements are shown as black crosses to indicate the astrometric errors; the insets are magnified by a factor of ten.}
\label{fig_orb_di}
\end{figure}

\begin{figure*}
\centering \includegraphics[width=0.79\textwidth]{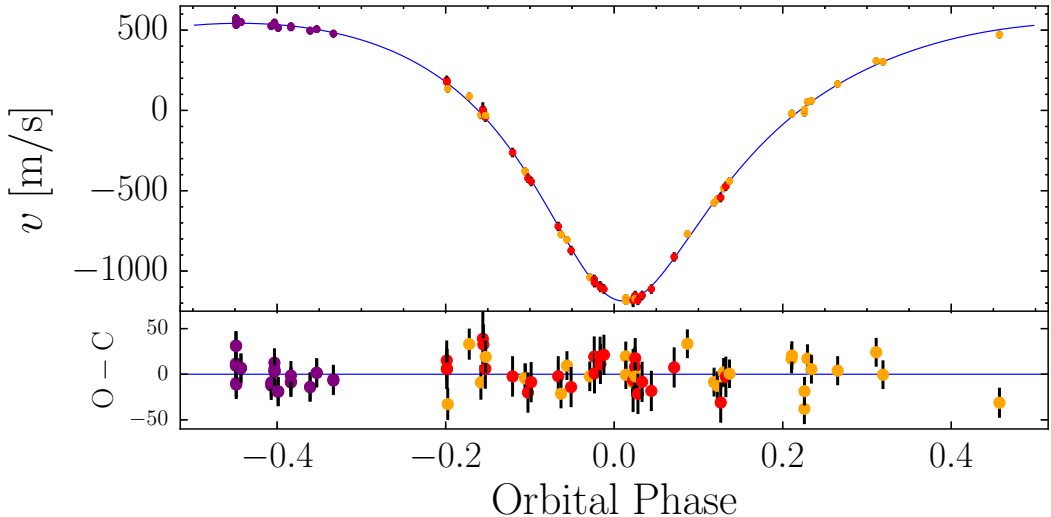}
\caption{Radial velocity measurements from ELODIE (red), Hamilton (orange) and
SOPHIE (purple) observations, and the fitted RV curve of V450~And~A, phase-folded
with the orbital period. Measurements are shifted by the values of offsets given in Table
\ref{tab_orb}. The lower panel shows the residuals of the fit.
The jitter is included in the error bars; they are plotted in the upper panel as well, but cannot be seen there as they are comparable to the size of the points.}
\label{fig_orb_rv}
\end{figure*}

\begin{figure*}
\centering\includegraphics[width=0.92\textwidth]{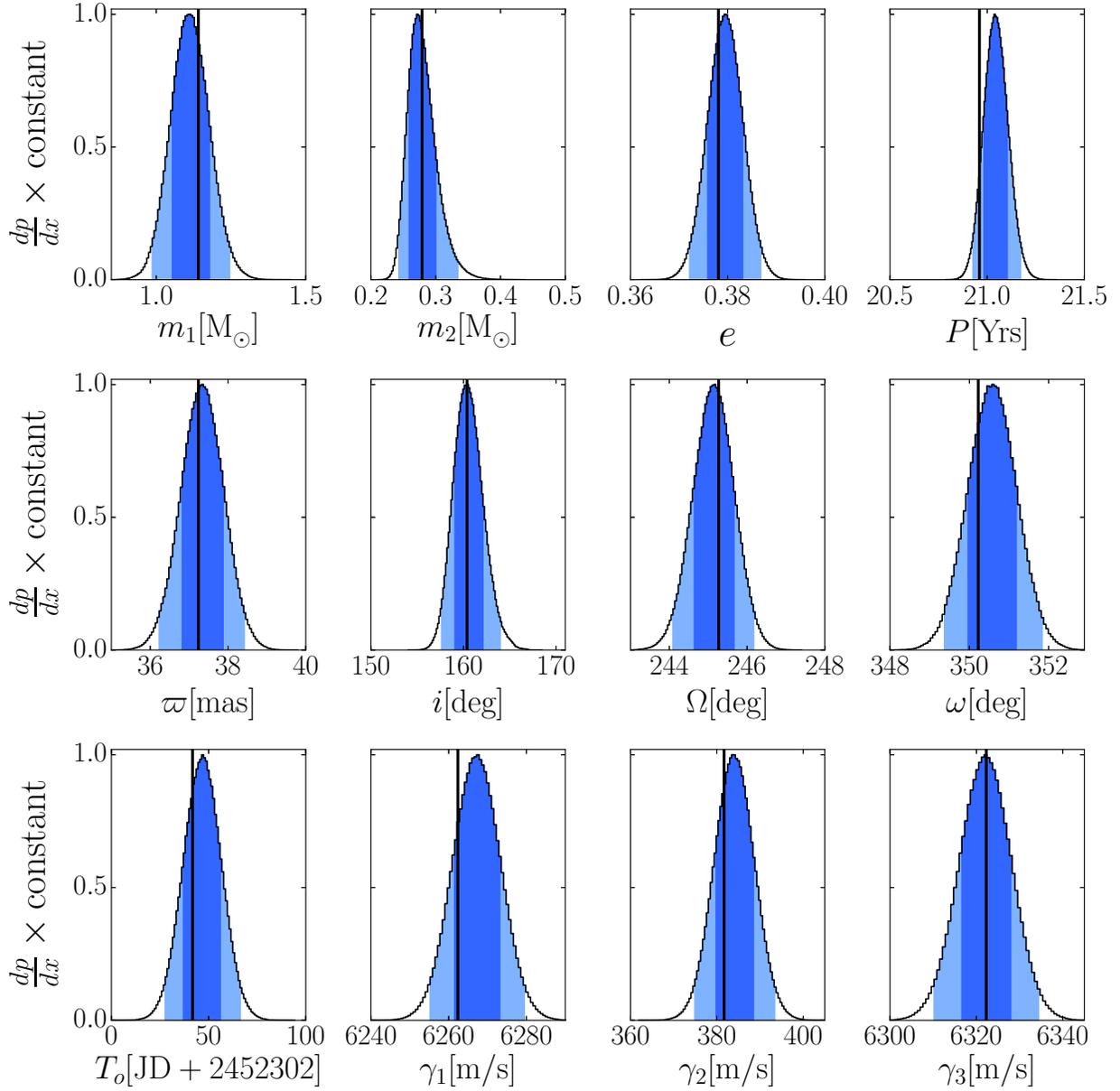}
\caption{The MCMC posterior distributions of the varying parameters. Best-fit values are 
marked with black vertical lines, while the dark and light blue ranges represent
the 68.3 and 95.4\% (1 and 2$\sigma$) confidence levels, respectively. 
The consecutive offsets ($\gamma$) are for the ELODIE, Hamilton, and SOPHIE instruments.
}\label{fig_mcmc}
\end{figure*}

\begin{figure}
\centering\includegraphics[width=\columnwidth]{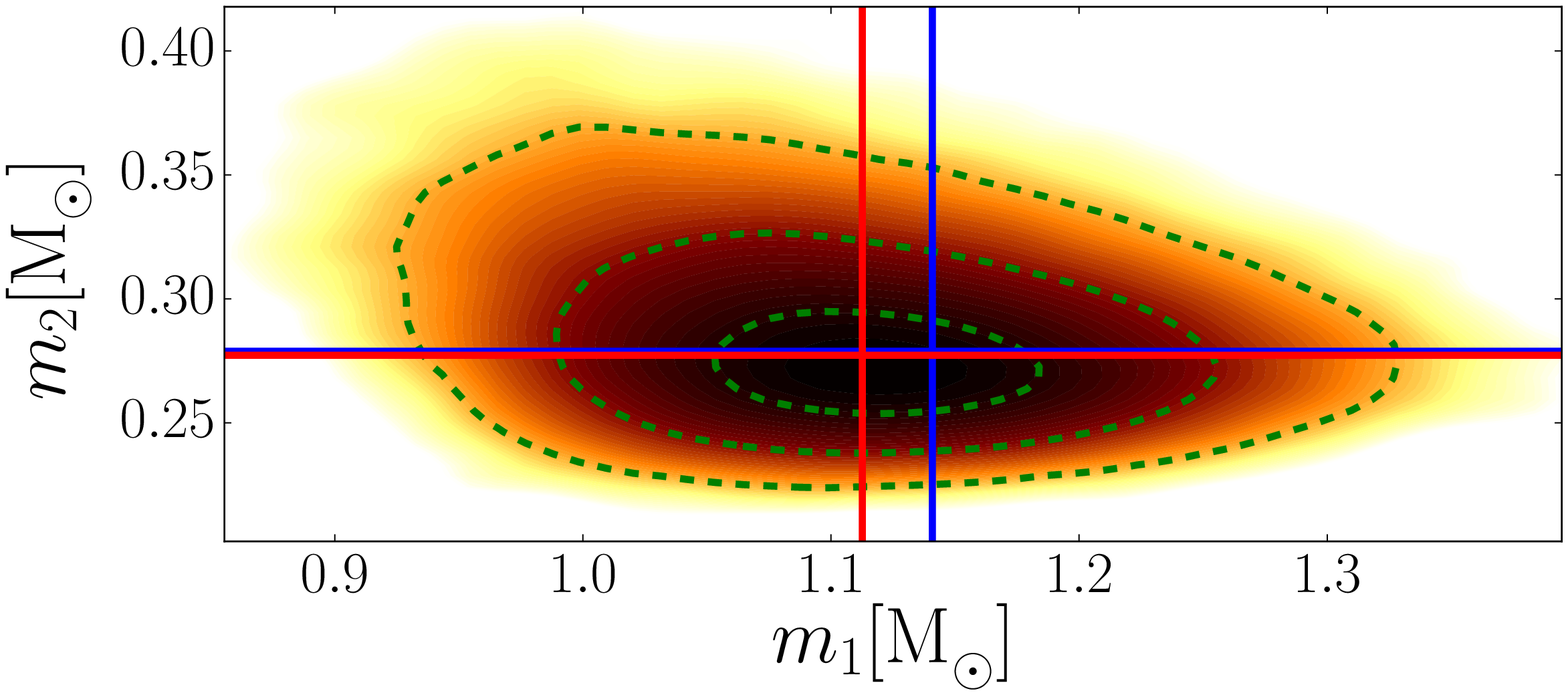}
\caption{A 2-dimensional density plot for comparison of the final posterior density function for $m_1$ and $m_2$. Dashed contours indicate the $3\sigma$, $2\sigma$, and $1\sigma$ confidence levels, with the best fit and median values as solid blue and red lines.}\label{fig_m1m2dens}
\end{figure}

Prior to utilizing both forms of observational data, initial fits were performed with only the radial velocity data. During this, an instrument-independent jitter was added in quadrature with the estimated RV uncertainties, and adjusted until a value of 13.6 \ms  produced a best reduced $\chi^2=1$. Following this step, a full joint analysis was performed to find the posterior probability distributions of the orbital parameters. Figures \ref{fig_orb_di} and \ref{fig_orb_rv} show the best fit orbit in both astrometry and RV. Table \ref{tab_orb} summarizes both the best-fitting and median values, as they differ due to asymmetries in the posterior distributions. For the MCMC stage we ran 7 parallel chains of 2$\times 10^8$ samples each to achieve sufficient convergence to the posteriors, shown in Figure \ref{fig_mcmc}, measured by a Gelman-Rubin statistic value of 1.0006.

We found that the secondary is a 0.279$^{+0.023}_{-0.020}$~M$\odot$ star (suggesting it is an M dwarf) on a nearly face-on, eccentric orbit. The orbital period of 21 years is over two times longer than suggested in P03, which is mainly due to the much shorter time span of their data ($\sim$4.5~years), which led to a poor determination of orbital parameters. The resulting $m_2\sin(i)$ is therefore much larger than in P03, and the companion cannot be a brown dwarf.
We were a bit fortunate that our imaging covered the apocenter passage. The orbital parameters are given with a very good precision (e.g.: 0.53\% in $K$), mainly thanks to the almost complete (87\%) coverage of the orbit with RV data, despite the stellar activity. The near face-on orientation of the orbit adds to the difficulty in determining the companion mass, therefore our 0.277$_{-0.019}^{+0.024}$~M$_\odot$ median fit for $m_2$ has $\sim$10\% error. These values arise from our choice to make both the primary star's mass and the parallax directly varied parameters, ensuring their errors are appropriately handled at every step in the MCMC chains and the resulting posterior distributions.

As shown in Figure \ref{fig_m1m2dens}, the posterior probability distributions for the masses $m_1$ and $m_2$ are anti-correlated.
This is largely due to Kepler's third law:
\begin{equation}
a_{\rm total} = \bigg[\frac{P^2G(m_{\rm total})}{4\pi^2} \bigg]^{1/3}
\label{eq_kepThird}
\end{equation}
As listed in Table \ref{tab_orb}, both the $a_{\rm total}$ and $P$ fits are at least $\sim$4 times tighter than for the masses. Equation \eqref{eq_kepThird} requires that the total mass ($m_{\rm total}=m_1+m_2$) must also have a relatively narrow distribution, even after taking the cube of $a_{\rm total}$. 

\begin{deluxetable}{lccc}
\tablewidth{0pt}
\tablecaption{Results of the joint astrometric+RV Keplerian orbital fit.}
\tablehead{
    \colhead{Parameter} &
    \colhead{Best-fit} &
    \colhead{Median} &
    \colhead{68\% range\tablenotemark{a}}}
\startdata
$m_1$ (M$_\odot$)     & 1.141 & 1.113 & (1.049:1.178) \\
$m_2$ (M$_\odot$)     & 0.279 & 0.277 & (0.259:0.301) \\
$\varpi$ (mas)        & 37.24 & 37.34 & (36.79:37.89) \\
$e$ (\ )              & 0.3781 & 0.3795 & (0.3759:0.3831) \\
$T_0$ (JD-2452302)    & 41.7  & 46.8  & (37.0:56.5) \\
$P$ (yr)              & 20.960 & 20.044 & (20.983:21.109) \\
$i$ ($^\circ$)        & 160.4 & 160.5 & (159.0:162.2) \\
$\Omega_2$ ($^\circ$) & 245.27 & 245.14 & (244.61:245.67) \\
$\omega_2$ ($^\circ$) & 350.22 & 350.58 & (349.96:351.19) \\
$a_{total}$ (AU)      & 8.54  & 8.51  & (8.38:8.64)\\
$K$ (\ms)             & 864.8 & 865.5 & (861.0:870.1) \\
$\gamma_{_{\rm{ELO}}}$ (\ms)\tablenotemark{b}  & 6262 & 6267 & (6261:6273) \\
$\gamma_{_{\rm{Ham}}}$ (\ms)\tablenotemark{b}  &  381.7 &  384.0 & (379.4:389.7) \\
$\gamma_{_{\rm{SOP}}}$ (\ms)\tablenotemark{b}  & 6322 & 6322 & (6316:6328) \\
$\chi^2$ $(\nu)$\tablenotemark{c}      & 83.87 (76) & \nodata  & \nodata 
\enddata
\tablenotetext{a}{Approximate 1$\sigma$ uncertainty: 68.3\% of the posterior probability lies in this range.}
\tablenotetext{b}{Radial velocity offsets: systemic velocities for ELODIE and SOPHIE, difference between the average and systemic velocity for Hamilton.}
\tablenotetext{c}{$\nu$ represents the number of degrees of freedom during MCMC joint fitting $\nu$=(4 DI epochs)(2 dimensions)+(80 RV epochs)(1 dimension)-(12 varied params)=88-12=76 dof.}
\label{tab_orb}
\end{deluxetable}

\section{Spectral analysis}\label{sec_spectr}

The literature values of atmospheric parameters of V450~And, such as the effective temperature
$T_{eff}$, metallicity $[M/H]$, or logarithm of gravity $\log(g)$, are not
always in agreement with each other. Temperatures vary from 5546 \citep{hol09}
to 5755~K \citep{val05}, $\log{g}$ from 4.35 \citep{gra03} to 4.60 \citep{tak07},
and the metallicity from -0.30 \citep{hol09} up to +0.07~dex \citep{kov04}.
As our intention was to compare our results with theoretical models and obtain as complete description of the system as possible, we decided to perform our own spectral analysis, based on the SOPHIE spectra. We used the freely available, python-based software iSpec 
\citep{bla14}\footnote{https://www.blancocuaresma.com/s/iSpec/}, that
compares an observed spectrum with a number of synthetic ones generated on-the-fly, and uses
a least-squares minimization algorithm. We first fit the continuum, but we cropped 
all the spectra at 4000~\AA, because the wide calcium H and K lines, and poor SNR
in the blue part, make the
continuum fitting difficult for shorter wavelengths. No errors were assigned to
the (relative) fluxes. We also corrected all the spectra for their barycentric 
velocity. We set the spectral resolution to 75000 (as in the instrument's 
specifications), and fitted for the temperature, metallicity, $\log(g)$ and 
projected rotational velocity $v \sin(i)$. Following \citet{val05}, we set 
the microturbulence velocity $v_{mic}$ to 0.85~k\ms, and for the macroturbulence 
velocity $v_{mac}$ we adopted their Equation (1), assuming $T_{eff}=5700$~K, 
which gives $v_{mac}=3.87$~k\ms. The relation predicts a change in $v_{mac}$ by 
1 k\ms\ per 650~K, so we conclude that the systematic uncertainty of the fitted 
parameters, coming from incorrect temperature we assumed, is negligible. We performed a number of tests, running iSpec on spectra of several stars with well-constrained parameters, provided together with the iSpec itself, or by the {\it Gaia}-ESO survey. We found that the consistency with the reference values is the best when iSpec uses the MARCS~GES atmosphere models \citep{gus08}, solar abundances from \citet{gre07}, and VALD atomic lines \citep{kup11}. \par
The iSpec was run on each SOPHIE spectrum separately. Individual results derived from each spectrum were then weight-averaged, and these values are given in Table \ref{tab_ispec}. They are all within the ranges of values found in the literature. The errors for the weight-averaged results are indicated by ``$\pm$'', and the $rms$ of all individual results are also provided in Table \ref{tab_ispec}. Errors of the individual iSpec results were larger than the $rms$. For each parameter we also give the mean of the individual errors.

\begin{deluxetable}{lcccc}
\tablewidth{0pt}
\tablecaption{Results of the iSpec spectral analysis}

\tablehead{
    \colhead{Parameter} &
    \colhead{Weigh.Avg.} &
    \colhead{$\pm$} &
    \colhead{\it rms} &
    \colhead{Mean ind.err.}}
    
\startdata
$T_{\rm eff}$~(K) & 5721 & 5 & 27 & 107\\
$\log(g)$ (cm s$^{-2}$)& 4.44 & 0.02 & 0.07 & 0.16\\
$[M/H]$ (dex)& 0.038 & 0.005 & 0.027 & 0.067\\
$v\sin(i)$~(k\ms) & 4.02 & 0.07 & 0.34 & 0.55
\enddata

\label{tab_ispec}
\end{deluxetable}

\section{Discussion}\label{sec_disc}
\subsection{Activity and Rotation}\label{sec_act}
\begin{figure}
\centering\includegraphics[width=\columnwidth]{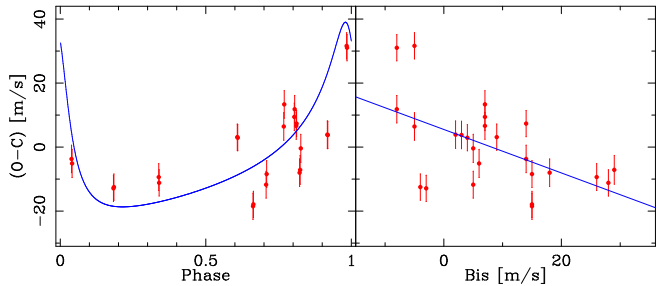}
\caption{{\it Left:} SOPHIE residuals of the orbital fit phase-folded with the period of rotation. The blue line shows a pseudo-orbital fit to these residuals, that mimics a 0.14~M$_J$ planet on an $e=0.62$ orbit. {\it Right:} The same residuals shown as a function of bisector span. The blue line shows a linear fit, and the 
correlation coefficient is -0.602. The vertical error bars on both panels are the RV measurement errors that we used in our analysis before adding the jitter.}
\label{fig_rvres}
\end{figure}

As a BY~Dra type variable, V450~And has been known to show some level of activity. Emission in Ca {\sc ii} H and K lines has been noted, 
\citep{str00,gra03,wri04,isa10}, and a photometric variability of 0.02~mag
has also been reported \citep[][]{str00}. In Section \ref{sec_hip_phot} we described
how the period of rotation (5.743~d) has been found. The same period can be seen 
in SOPHIE radial velocities, i.e. in the residuals $(O-C)$ of the orbital fit. 
In Figure \ref{fig_rvres} we present them as a function of the
rotation phase, and the bisector span $bis$. The correlation between $bis$ and 
$(O-C)$ proves the activity origin of the observed RV modulation. The activity of
V450~And~A mimics a 0.14~M$_J$ planet on an eccentric orbit -- a pseudo-orbital
fit to all the residuals is also shown, with an $rms$ of only $\sim$8~\ms. 
We would like to note that we
did not model the activity-related RV variations together with the orbital motion due
to two reasons. First, only the SOPHIE data are precise enough to see it clearly, 
and they do not cover a substantial part of the orbit. Second, we suspect that 
the pattern of spots (and thus the RV modulation they produce) evolves in time.
When making the pseudo-orbital fit to residuals from only one season, one can lower
the $rms$ to 2-3~\ms, and for each season obtain different parameters (shape) of the pseudo-orbit (different $e, K, \omega$, etc.).

In the further analysis we thus assumed 5.743~d to be the period of rotation.
Taking the best-fitting dynamical mass of the star, and $\log(g)$ from the 
spectral analysis, we can estimate the radius of V450~And~A to be 
1.09$^{+0.09}_{-0.11}$~R$_\odot$. This value of radius, the estimated rotational
period, and projected velocity of rotation $v\sin(i)$ lead to the rotation 
inclination angle of 24.8 or 155.2$\pm$5.5$^\circ$, which is in agreement with the
orbital inclination. Due to a long orbital period and young age (see next
Section), we posit that the system formed in this way, rather than was aligned 
by tidal forces. A nearly pole-on orientation of the component A may also
partially explain why the amplitude of photometric variability is only 0.02~mag --
if spots are present near the star's pole, their visibility does not change much
with the phase of rotation. Such polar or high-latitude spots have been observed 
on stars of similar mass, spectral type and rotation period \citep[e.g.,][]{str98,san13}. 
In addition, theoretical models support this tendency \cite[e.g.,][]{gra00,yad14}. 
Hence, it is likely that we observed spots near the star's pole, systematically 
affecting our luminosity and temperature estimates for V450~And~A (see 
Section \ref{sec_iso}). 

\subsection{Age from Activity and Gyrochronology}\label{sec_age}

\begin{figure}
\centering\includegraphics[width=\columnwidth]{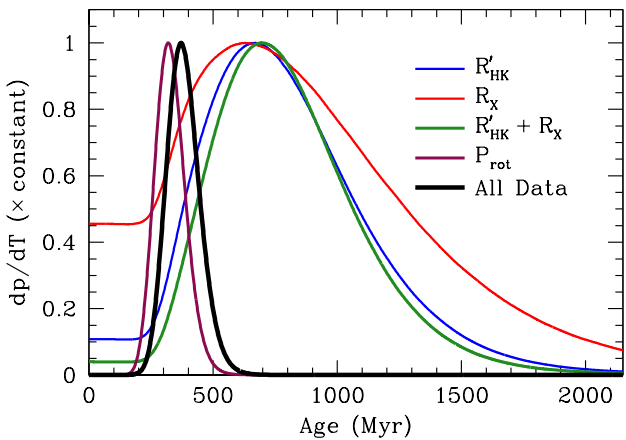}
\caption{Posterior probability distributions of the age of V450~And from its rotation and activity indicators, using the method described in \cite{bra14}. There is a modest tension between the age estimated from chromospheric activity and that inferred from rotation; our rotation period dominates the total posterior probability distribution.}
\label{fig_age}
\end{figure}

Stars with convective envelopes generate magnetic fields that couple them to their stellar winds. The stars shed angular momentum, spinning down and becoming less active with time \citep{sku72}. This spindown, either measured directly through a rotation period or indirectly through coronal and chromospheric activity, can be used to date cool main sequence stars \citep{bar03}. The decline of rotation and activity has been calibrated using clusters and binaries to enable gyrochronology \citep{bar07,mam08}.

V450 And A's $B-V$ was measured by Tycho-2 \citep{hog00} and we convert that to the Johnson $B-V$ ($= 0.690 \pm 0.015$), which is adopted in the following analysis.  V450 And has well-measured rotation and activity indicators, including the period of 5.7 days derived in the preceding section, a {\it ROSAT} X-ray flux \citep{vog99}, and Ca\,{\sc ii} HK chromospheric activity measurements. We use the bolometric corrections of \cite{flo96} as corrected by \cite{tor10} to convert the measured X-ray flux into $R_X$, the ratio of X-ray to bolometric flux; we obtain $\log_{10} R_X = -4.65$. From \cite{pac13}, there are 29 literature measurements of the Ca\,{\sc ii} HK S index, ranging from 0.235 to 0.348. The lowest measured value, from \cite{whi07}, is a clear outlier, while the single measurement from \cite{str00} must be calibrated to the Mount Wilson system. The remaining 27 measurements, from \cite{gra03}, \cite{wri04}, and \cite{isa10}, have a mean of 0.324 and a standard deviation of 0.013. The median of all literature measurements is an indistinguishable 0.322; we adopt this value as our Ca\,{\sc ii} HK index and convert it to $\log_{10} R'_{\rm HK} = -4.48$ using the formulae given in \cite{noy84}.

We adopt the approach of \cite{bra14} to obtain a posterior probability distribution on the age of V450 And from its rotation and activity. \cite{bra14} used the calibrated relations of \cite{mam08}, but also accounted for the fact that stars spend a variable amount of time as rapid rotators before they begin to spin down in earnest \citep{bar07}. For a $\sim$1 $M_\odot$ star, \cite{bra14} assumed an additional spin-down time uniformly distributed between 0 and $\sim$150 Myr, effectively broadening the posterior probability distributions.

Figure \ref{fig_age} shows the ages as inferred from the method of \cite{bra14} using the rotation and activity indicators individually, and by combining all age indicators. We obtain a final age estimate of 380$^{+220}_{-100}$ Myr, (at 90\% confidence). There is a mild tension between the age preferred by chromospheric activity (300--1800 Myr) and that favored by our measured rotation period, which dominates our posterior probability distribution. Following \cite{bra14}, we include a 5\% probability that the star's rotation and activity do not reflect its true age and that the star could be at any point on its main sequence evolution.

\subsection{Comparison with Models}\label{sec_iso}

\begin{deluxetable}{ccc}
\tablewidth{0pt}
\tablecaption{Absolute $JHK$ magnitudes of V450 And}

\tablehead{
    \colhead{} &
    \colhead{Primary} &
    \colhead{Secondary}}
    
\startdata
$M_J$~(mag) & 3.822 $\pm$ 0.037 & 8.002 $\pm$ 0.086 \\
$M_H$~(mag) & 3.530 $\pm$ 0.041 & 7.581 $\pm$ 0.044 \\
$M_{K_{\rm{s}}}$~(mag) & 3.455 $\pm$ 0.039 & 7.275 $\pm$ 0.088 
\enddata

\label{tab_absmag}
\end{deluxetable} 

\begin{figure*}
\centering\includegraphics[width=0.8\textwidth]{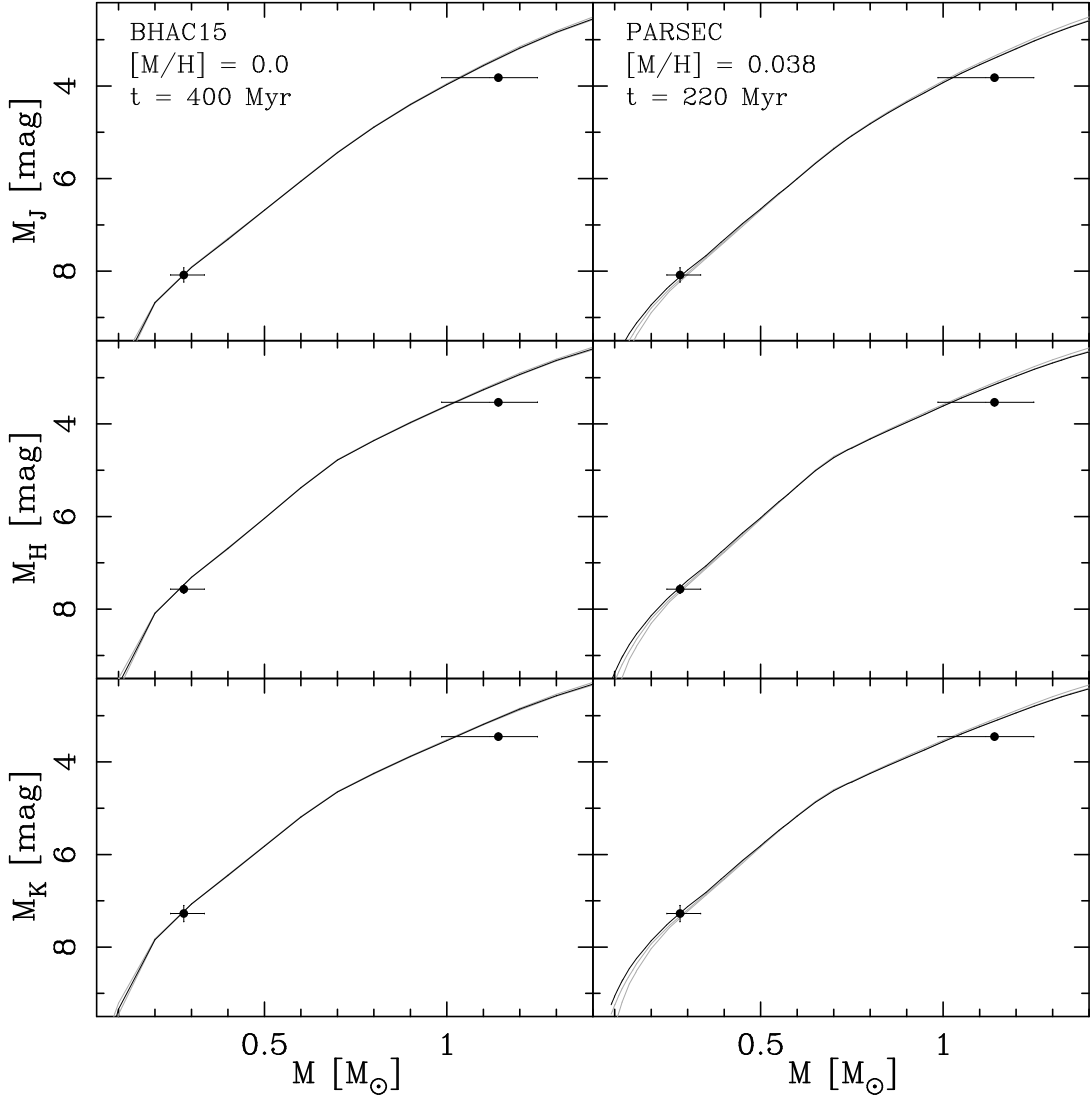}
\caption{Comparison of our results, with 95\% confidence level error bars, 
with theoretical models. 
The best-fitting isochrones of BHAC15 (400 Myr; left) and PARSEC (220 Myr; right)
sets are shown as black lines ($J$ [top], $H$ [middle], $K$-band [bottom])
on mass vs. absolute magnitude planes.
Grey lines are isochrones for ages of 280 and 600 Myr, which are 90\% 
confidence limits from Section \ref{sec_age}. Assumed metallicities are labeled for each set.
}\label{fig_iso}
\end{figure*}

In order to estimate the age of the system on the basis of theoretical stellar
evolution models, we compared our results with the \citet[][hereafter BHAC15]{bar15}
and PAdova-TRieste Stellar Evolution Code \citep[PARSEC;][]{bre12,che14} isochrones.
The former set is calculated only for the solar composition, but our iSpec analysis
gives metallicity very similar to solar. For PARSEC we used our value of [$M/H$], which 
translates into $Z=0.0166$ in this set. We checked and confirmed that using the solar
composition ($Z=0.0152$) does not change the final results significantly, so the usage of
[$M/H$]=0.0 BHAC15 models is justified.

For the comparison we use our estimates of dynamical masses and absolute $JHK_{\rm{s}}$
magnitudes, as they are the most robust parameters we can obtain from our observations
for both components. To compute the absolute magnitudes, we assumed the {\it Hipparcos}
distance to the system, and took the total apparent brightnesses of the system and our 
magnitude differences from HiCIAO. We list the absolute magnitudes in Table \ref{tab_absmag}.
The results are shown in Figure~\ref{fig_iso}. Errors of the values for both 
components come mainly from the parallax uncertainty, with a small contribution 
from the uncertainties of the observed magnitudes. For the secondary only, the 
second dominant source of errors is the differential photometry from 
Table~\ref{tab:obslog}.

Both sets of models give quite consistent results. The resulting ages are
$400^{+2600}_{-250}$~Myr ($\log(\tau)=8.60^{+0.88}_{-0.43}$) for BHAC15, and
$220^{+2120}_{-90}$~Myr ($\log(\tau)=8.34^{+1.03}_{-0.22}$) for PARSEC.
We adopt the latter because of the smaller error bars. 
The best-fitting age values are tightly constrained by
the best-fitting value of $m_2$, as found in the orbital fit. The lower limits
are defined by the 95\% confidence level of the parameters of the secondary, while the primary
analogously defines the upper limits. In other words, isochrones of the given range 
of ages reside within mass vs. absolute ($JHK_{\rm{s}}$) magnitude planes that are 
defined by 95\% confidence levels (corresponding to $\sim$2$\sigma$ for symmetrical
posteriors), as derived from both components. The large positive uncertainty is 
strictly related to the fact that stars do not change much during the main sequence 
phase, and isochrones for much older years still reside within the error bars shown in
Fig.~\ref{fig_iso}.

The primary tends to be systematically too faint for its mass. The PARSEC model, 
for $\tau=220$~Myr and [$M/H$] = 0.038, predicts absolute magnitudes lower 
by 0.92 in $U$, 0.65 in $V$, and 0.30~mag in $K$ with respect to observed values -- 
5.94 \citep{mye15}, 5.74 \citep{hog00}, and 3.46~mag \citep{cut03}, respectively, 
as derived from literature data and distance. This is mainly because
the effective temperature at the age of $\sim$220~Myr is $\sim$450~K higher in 
models than what we found with iSpec. This can be at least partially explained by 
the presence of polar spots, seen in photometry, which decrease
the observed $T_{eff}$ and brightness. 

Our analysis also shows that the secondary has 
just reached the main sequence, but spots may also hamper its results, making it appear 
fainter. In such a case the true absolute magnitude would be different, and the secondary 
would move up on Fig. \ref{fig_iso}.

Ages derived from comparison with both sets of models are in good 
agreement with results from Section \ref{sec_age}. In Fig.~\ref{fig_iso} we also plot 
isochrones for ages of 280 and 600 Myr, which are the 90\% confidence limits of the
\citet{bra14} method. With a bit higher temperature of the primary, the agreement would be 
even better. 
Additionally, our $m_1$ determination may be slightly overestimated, due to uncertainties in parallax or astrometry, for example. Please also note that the median of the $m_1$ posterior distribution is smaller. Because, as discussed in Section \ref{sec_orbSoln}, the sum of two masses is fairly well constrained, this would mean that $m_2$ is higher. Including the possible effect of spots on component B, described above, the secondary would then move on Fig.~\ref{fig_iso} almost exactly along the isochrone. The resulting age would thus be intact. This issue would be fixed if radial velocities of the secondary were measured directly.

Overall, we obtained a consistent image of the V450~And binary. It is a fairly young, few hundred Myr old system, but with both components already on the main sequence. Probably both stars are active (primary is for sure), and possibly their mass ratio is slightly closer to 1, as compared with the best-fitting value we obtained. Our age estimates, especially from Sect.~\ref{sec_age}, agree well with one of the two mostly accepted determinations of the age of the Castor moving group: $440\pm40$~Myr \citep{mam12}. The less-agreeing one is $700^{+150}_{-75}$~Myr \citep{mon12}. However, it was recently suggested by \citet{mam13} and \citet{zuc13} that this moving group is actually a collection of kinematically similar stars showing a spread of ages, rather than a well-defined kinematic structure. 

\section{Summary and future prospect}\label{sec_sum}

In this work we have presented our analysis of a low-mass companion originally discovered 
with the RV technique, and since detected with high-contrast imaging. In comparison with 
previous work by P03, we employ in our analysis additional astrometry and more RV data,
covering almost the whole orbital period. This enables us to better constrain the 
orbit, as well as other crucial systems parameters. We derived the full set of orbital
parameters and masses for both components, showing that the secondary is an M-type star, 
not a brown dwarf as suggested by P03. With the support of the results of our spectral 
analysis, and the observed activity of the primary, we conclude that the system is 
still relatively young ($\sim$200-400~Myr). The combined RV and imaging data allowed us 
to draw a full image of the V450~And system.

The presented work may be a test case for future studies, aimed for characterisation of 
known brown dwarf and exoplanet candidates. Cases like V450~And, HD~16760, or
objects from the recent study by \citet{wil16}, clearly show the need to support RVs 
with other kinds of data. With the capabilities of already existing 
extreme adaptive optics systems, like SCExAO at Subaru \citep{mar14,jov15}, GPI at Gemini 
\citep{mac14} or SPHERE at VLT \citep{beu08}, and their future generations that will be
working with the incoming 30-m class telescopes, it will be possible to detect and directly
characterise companions of lower masses and shorter orbital periods than V450~And~B. 
This may bring a revision of our knowledge of long-period brown dwarf and massive planet
candidates, their distribution, initial mass function, and mechanisms of their 
formation, for example by pointing out objects in the ``brown dwarf desert'' 
\citep{mar00,gre06,wil16}.

\acknowledgments

We would like to thank Dr. Rodolfo Smiljanic and Dr. Milena Ratajczak, both from the N.~Copernicus Astronomical Center, Poland, for providing the {\it Gaia}-ESO survey spectra for tests, and for their comments regarding the spectral analysis. We are also grateful to Dr. Roberto Sanchis-Ojeda for a useful discussion, and to David Lafreni\'{e}re for generously providing the source code for the ADI data reductions. 

This work is based in part on data collected at Subaru Telescope, which is operated by the National Astronomical Observatory of Japan. Our data reductions in part were performed with PyRAF and PyFITS, which are products of the Space Telescope Science Institute (STScI/AURA/NASA). Our analysis is also based on observations made with the NASA/ESA Hubble Space Telescope, and obtained from the Hubble Legacy Archive, which is a collaboration between the Space Telescope Science Institute, the Space Telescope European Coordinating Facility (ST-ECF/ESA) and the Canadian Astronomy Data Centre (CADC/NRC/CSA). Data analysis were in part carried out on common use data analysis computer system at the Astronomy Data Center, ADC, of the National Astronomical Observatory of Japan. This research has made use of the VizieR catalogue access tool and the SIMBAD database, which are operated at CDS, Strasbourg, France. 

K.G.H. acknowledges support provided by the National Astronomical Observatory of Japan as Subaru Astronomical Research Fellow. M.K. was supported by Japan Society for Promotion of Science (JSPS) Fellowship for Research and this work was partially supported by the Grant-in-Aid for JSPS Fellows (Grant Number 25-8826). K.M. gratefully acknowledges support from the Mitsubishi Corporation International Student Scholarship. This work was performed in part under contract with the Jet Propulsion Laboratory (JPL) funded by NASA through the Sagan Fellowship Program executed by the NASA Exoplanet Science Institute. N.N. acknowledges supports by the NAOJ Fellowship, Inoue Science Research Award, and Grant-in-Aid for Scientific Research (A) (JSPS KAKENHI Grant Number 25247026). J.C.C. was supported by the U.S. National Science Foundation under Award No. 1009203. M.T. is supported by JSPS KAKENHI Grant (Number 15H02063).

Finally, the authors recognize and acknowledge the very significant cultural role and reverence that the summit of Maunakea has always had within the indigenous Hawaiian community. We are most fortunate to have the opportunity to conduct observations from this mountain.

{\it Facilities:} \facility{Subaru (HiCIAO, AO188)}, \facility{OHP:1.93m (ELODIE, SOPHIE)}, \facility{Shane (Hamilton spectrograph)}

\appendix

In Table \ref{tab_rvall} we present all RV measurements used in this study, given in the original form and numerical precision as in the source. For each measurement we also give the final measurement error (including systematics but without the jitter), the value of RV with respect to the offset (RV-$\gamma$), residual (O-C), and the instrument that was used.

\LongTables
\begin{deluxetable}{rcrrc}
\tablewidth{0pt}
\tablecaption{Radial velocities of V450~And\label{tab_rvall}}
\tablehead{
    \colhead{BJD} &
    \colhead{RV $\pm$ err} &
    \colhead{RV-$\gamma$} &
    \colhead{O-C} &
    \colhead{Sp.\tablenotemark{a}} \\
    -2450000 & (\ms) & (\ms)  & (\ms) & 
    }
\startdata
 822.3353812 &  6380  $\pm$ 15.1 &   184.0 &  12.6 & E \\
 823.3797985 &  6370  $\pm$ 15.1 &   174.0 &   3.2 & E \\
 824.3384884 &  6370  $\pm$ 15.1 &   174.0 &   3.7 & E \\
 831.6181600 & 516.60 $\pm$ 8.12 &   134.5 & -32.1 & L \\
1027.0039100 & 468.65 $\pm$ 7.30 &    86.5 &  33.9 & L \\
1132.8232400 & 354.28 $\pm$ 10.76&   -27.8 &  -8.0 & L \\
1150.3587309 &  6170  $\pm$ 30.8 &     4.0 &  36.6 & E \\
1155.4064768 &  6190  $\pm$ 15.5 &    -6.0 &  30.4 & E \\
1173.3595108 &  6150  $\pm$ 15.1 &   -46.0 &   3.8 & E \\
1175.7724600 & 350.44 $\pm$ 8.73 &   -31.7 &  19.9 & L \\
1420.6268456 &  5930  $\pm$ 15.4 &  -266.0 &  -4.8 & E \\
1533.8017600 &  3.94  $\pm$ 4.16 &  -378.2 &  -3.1 & L \\
1536.6826200 &  0.00  $\pm$ 4.09 &  -382.1 &  -4.0 & L \\
1560.3601856 &  5770  $\pm$ 15.2 &  -426.0 & -22.7 & E \\
1589.2844593 &  5750  $\pm$ 15.2 &  -446.0 & -11.3 & E \\
1835.5565444 &  5470  $\pm$ 15.1 &  -726.0 &  -5.5 & E \\
1860.8232400 &-390.08 $\pm$ 4.72 &  -772.2 & -21.5 & L \\
1914.6845700 &-423.26 $\pm$ 3.96 &  -805.4 &   9.1 & L \\
1952.3257010 &  5320  $\pm$ 15.2 &  -876.0 & -17.7 & E \\
2121.9785200 &-657.11 $\pm$ 5.36 & -1039.2 &  -2.9 & L \\
2162.5770622 &  5140  $\pm$ 15.4 & -1056.0 &  15.5 & E \\
2164.6197949 &  5120  $\pm$ 15.2 & -1076.0 &  -2.8 & E \\
2215.4971564 &  5100  $\pm$ 15.2 & -1096.0 &  15.4 & E \\
2218.5181787 &  5090  $\pm$ 15.1 & -1106.0 &   7.4 & E \\
2250.3889077 &  5080  $\pm$ 15.3 & -1116.0 &  17.5 & E \\
2448.9902300 &-785.27 $\pm$ 5.19 & -1167.4 &  19.7 & L \\
2449.9845300 &-805.42 $\pm$ 3.24 & -1187.5 &  -0.6 & L \\
2509.9082000 &-796.22 $\pm$ 2.98 & -1178.3 &  -0.6 & L \\
2512.0263700 &-800.46 $\pm$ 3.48 & -1182.6 &  -5.3 & L \\
2513.0146500 &-803.54 $\pm$ 3.24 & -1185.7 &  -8.7 & L \\
2515.6315310 &  5010  $\pm$ 30.1 & -1186.0 &  -9.7 & E \\
2532.6263008 &  5030  $\pm$ 15.2 & -1166.0 &   5.4 & E \\
2534.6200153 &  5040  $\pm$ 15.3 & -1156.0 &  14.8 & E \\
2534.9335900 &-790.54 $\pm$ 3.77 & -1172.7 &  -2.0 & L \\
2559.6019814 &  5010  $\pm$ 15.3 & -1186.0 & -24.1 & E \\
2597.4354178 &  5040  $\pm$ 15.3 & -1156.0 & -10.9 & E \\
2681.3382643 &  5080  $\pm$ 15.3 & -1116.0 & -20.6 & E \\
2888.6594807 &  5280  $\pm$ 15.2 &  -916.0 &   5.3 & E \\
3008.7783200 &-387.63 $\pm$ 3.93 &  -769.7 &  34.3 & L \\
3251.0195300 &-193.93 $\pm$ 3.89 &  -576.0 &  -8.4 & L \\
3280.9150400 &-170.27 $\pm$ 3.91 &  -552.4 & -12.6 & L \\
3311.5240765 &  5650  $\pm$ 15.3 &  -546.0 & -34.2 & E \\
3337.6767600 &-104.14 $\pm$ 3.88 &  -486.3 &   1.9 & L \\
3358.3409466 &  5720  $\pm$ 15.5 &  -476.0 &  -6.2 & E \\
3391.6575900 & -58.13 $\pm$ 3.32 &  -440.2 &   0.3 & L \\
3954.9796600 & 357.90 $\pm$ 4.19 &   -24.2 &  15.2 & L \\
3957.9836700 & 363.23 $\pm$ 3.42 &   -18.9 &  18.8 & L \\
4071.8361900 & 366.27 $\pm$ 6.76 &   -15.8 & -39.8 & L \\
4072.7411200 & 386.03 $\pm$ 3.61 &     3.9 & -20.5 & L \\
4099.7956300 & 435.51 $\pm$ 3.71 &    53.4 &  15.2 & L \\
4135.6839700 & 441.76 $\pm$ 3.98 &    59.6 &   3.5 & L \\
4373.9683300 & 546.37 $\pm$ 4.43 &   164.3 &   1.6 & L \\
4723.9418500 & 690.14 $\pm$ 4.06 &   308.0 &  21.5 & L \\
4785.8204800 & 684.08 $\pm$ 3.82 &   302.0 &  -3.0 & L \\
5843.9587200 & 853.85 $\pm$ 6.00 &   471.7 & -33.4 & L \\
6562.4748310 & 6874.0 $\pm$ 4.3  &   552.9 &   9.4 & S \\
6562.4821578 & 6876.4 $\pm$ 4.3  &   555.3 &  11.8 & S \\
6563.5081427 & 6896.2 $\pm$ 4.1  &   575.1 &  31.6 & S \\
6563.5154927 & 6895.6 $\pm$ 4.1  &   574.5 &  31.0 & S \\
6565.5432908 & 6855.2 $\pm$ 4.2  &   534.1 &  -9.4 & S \\
6565.5506060 & 6853.4 $\pm$ 4.1  &   532.3 & -11.2 & S \\
6608.4598096 & 6871.2 $\pm$ 4.2  &   550.1 &   6.6 & S \\
6608.4677262 & 6871.9 $\pm$ 4.1  &   550.8 &   7.3 & S \\
6883.5463382 & 6846.3 $\pm$ 4.2  &   525.2 & -11.7 & S \\
6883.5563276 & 6849.6 $\pm$ 4.2  &   528.5 &  -8.4 & S \\
6912.6107890 & 6863.1 $\pm$ 4.3  &   542.0 &   6.4 & S \\
6912.6198869 & 6870.0 $\pm$ 4.3  &   548.9 &  13.3 & S \\
6913.4711873 & 6860.5 $\pm$ 4.3  &   539.4 &   3.9 & S \\
6913.4797411 & 6860.4 $\pm$ 4.3  &   539.3 &   3.8 & S \\
6946.4649120 & 6836.4 $\pm$ 4.1  &   515.3 & -18.5 & S \\
6946.4722502 & 6837.0 $\pm$ 4.1  &   515.9 & -17.9 & S \\
7062.2501790 & 6839.4 $\pm$ 4.3  &   518.3 &  -8.0 & S \\
7062.2613470 & 6840.3 $\pm$ 4.5  &   519.2 &  -7.1 & S \\
7062.2731630 & 6847.0 $\pm$ 4.4  &   525.9 &  -0.4 & S \\
7236.6208314 & 6818.8 $\pm$ 4.1  &   497.7 & -12.9 & S \\
7236.6298368 & 6819.2 $\pm$ 4.1  &   498.1 & -12.5 & S \\
7296.4982802 & 6828.1 $\pm$ 4.1  &   507.0 &   3.1 & S \\
7296.5072853 & 6827.9 $\pm$ 4.1  &   506.8 &   2.9 & S \\
7448.2976778 & 6801.0 $\pm$ 4.3  &   479.9 &  -3.7 & S \\
7448.3083946 & 6799.6 $\pm$ 4.4  &   478.5 &  -5.1 & S
\enddata
\tablenotetext{a}{``E'' stands for ELODIE, ``H'' for Hamilton, and ``S'' for SOPHIE spectrograph.}
\end{deluxetable} 

\bibliography{refs}

\end{document}